\documentclass[%
 reprint,
superscriptaddress,
 amsmath,amssymb,
 aps,
floatfix,
]{revtex4-2}

\usepackage{graphicx}
\usepackage{dcolumn}
\usepackage{bm}
\usepackage{color}
\usepackage{epstopdf}
\graphicspath{"/home/anshuman/Desktop/submission_Arxiv"}

\usepackage{xcolor,hyperref}
\hypersetup{
   colorlinks,
   linkcolor={blue!50!black},
   citecolor={blue!50!black},
   urlcolor={blue!80!black}
}

\begin{document}

\title{
Unconventional rheological properties in systems of deformable particles}

\author{Anshuman Pasupalak}
\author{Shawn Khuhan Samidurai}
\author{Yanwei Li}
\author{Yuanjian Zheng}
\affiliation{Division of Physics and Applied Physics, School of Physical and Mathematical Sciences\\
 Nanyang Technological University
}
\author{Ran Ni}
\affiliation{Division of Physics and Applied Physics, School of Physical and Mathematical Sciences\\
 Nanyang Technological University
}
\affiliation{School of Chemical and Biomedical Engineering\\
 Nanyang Technological University
}

\author{Massimo Pica Ciamarra}
\affiliation{Division of Physics and Applied Physics, School of Physical and Mathematical Sciences\\
 Nanyang Technological University
}
\affiliation{CNR–SPIN, Dipartimento di Scienze Fisiche, Università di Napoli Federico II, I-80126, Napoli, Italy}

\date{\today}

\begin{abstract}
We demonstrate the existence of unconventional rheological and memory properties in systems of soft-deformable particles whose energy depends on their shape, via numerical simulations.
At large strains, these systems experience an unconventional shear weakening transition characterized by an increase in the mechanical energy and a drastic drop in shear stress, which stems from the emergence of short-ranged tetratic order. 
In these weakened states, the contact network evolves reversibly under strain reversal, keeping memory of its initial state, while the microscopic dynamics is irreversible.
\end{abstract}

\maketitle
\section{Introduction}
When compressed to high density, soft particles such as microgels, foams and cells in tissue deform, develop facets and tessellate available space, provided they do not interpenetrate.
In these high-density systems, pair-wise interacting descriptions lose their value~\cite{likos2001effective} as the energy of a particle depends on the positions of all its neighbors and is defined through a many-body potential.
This potential is the work needed to squeeze the particle into its volume, and establishes a close link between the mechanical and the geometrical properties of the system. 
This link has recently been uncovered in a two-dimensional model, when the energy of a particle depends on its perimeter~\cite{Bi, Kim2018, li2018role, popovic2020inferring, Li2021}.
In this case, an elementary relaxation event involving the disappearance of an edge separating two particles of length $l$, has an energy cost $\propto l$.
Hence, the system’s mechanical response correlates with the edge-length distribution, $P(l)$, and in particular, the system loses mechanical rigidity when an extensive number of zero-length edges appear, $P(l)\propto l^\theta$ with $\theta = 0$~\cite{Bi, Kim2018, li2018role, popovic2020inferring, Li2021}.
When a particle's energy depends not only on its perimeter but also on other geometrical properties, such as the area, it is not clear how the intimate relationship between geometrical and mechanical properties manifests.

Here we show that the connection between energy and shape of a particle induces a novel shear weakening crossover and unusual memory properties in various models that differ in the shape dependence of the energy of the particles.
These models exhibit distinct rheological responses at small strains.
However, all of them weaken at large strains and their shear stress becomes negligible.
During this weakening crossover, particles acquire model-independent anisotropic shapes, orient analogously in space, and develop short-range tetratic order~\cite{donev2006tetratic,hou2020emergent,geng2009theory,Li2019b}.
In the strain weakened regime, all models exhibit unconventional memory properties in that under strain reversal, the microscopic dynamics evolves irreversibly, while the contact network changes reversibly.
These results demonstrate that the link between energy and geometry induces, in various models, a novel rheological response and unconventional memory properties.
\begin{figure*}[!!t]
	\centering
	\includegraphics[width=0.9\textwidth]{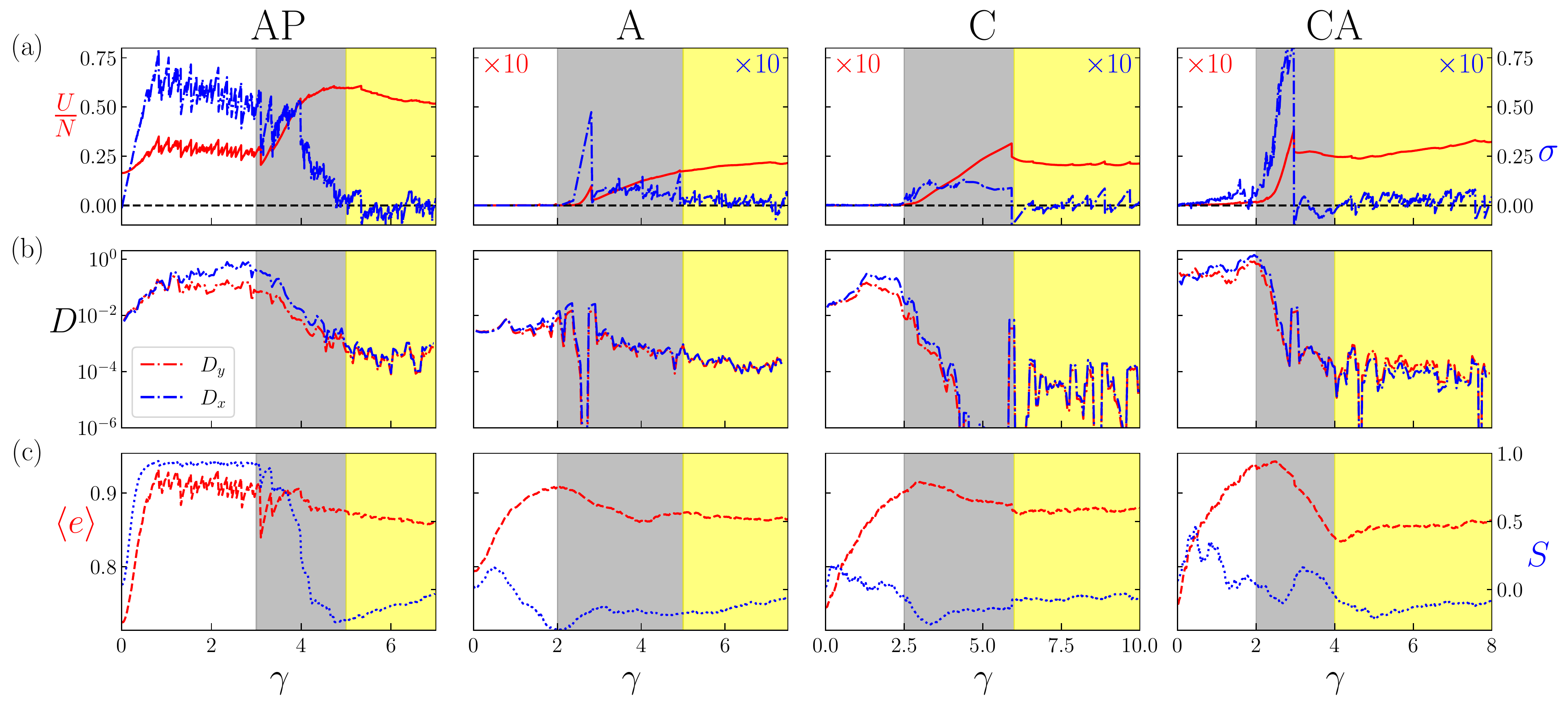}
	\caption{(a) Dependence of the mechanical energy (red) and shear stress (blue), (b) the transverse transverse (red) and longitudinal (blue) diffusivity, and (c) the eccentricity (red) and nematic order parameter (blue) on the shear strain, for all the models. 
		Gray and yellow shaded regions identify the strain weakening crossover, and the strain weakened regime.}
	\label{fig:energy_stress}
\end{figure*}

\section{Models and Methods}
We consider models suitable to describe systems of deformable particles compressed to high density that tessellate the space in two dimensions.
The centres of the particles ${\bf r}_i$ are the configurational degrees of freedom.
We associate a cell to each centre through a Voronoi tessellation, and assume the cells to correspond to the particles' shapes.
We numerically investigate four models describing systems of deformable particles at density $\rho=1$ at which the average particle area is $1$. 
The energy depends on how area $a_i$, perimeter, $p_i$ and center ${\bf r}_i$ of each particle respectively differ from the average area, a dimensionless perimeter $p_0$, and the centre of mass ${\bf r}_i^{(c)}$, 
\begin{equation}
U = \sum_{i=1}^{N} k_a(a_i-1)^2 + k_p(p_i-p_0)^2 + \epsilon ({\bf r}_i - {\bf r}_i^{(c)})^2,
\label{eq:energy_gen}
\end{equation}
with $k_a$, $k_p$ and $ \epsilon$ parameters.
We study (AP) the area+perimeter model, corresponding to the Voronoi model for epithelial cell tissue~\cite{bi2016motility, sussman2018no, li2018role,pasupalak2020hexatic,Zheng2020}, $k_a = 1$, $k_p = 1$, $\epsilon = 0$; 
(C) the centroid or Voronoi liquid model~\cite{Hain2020, Ruscher2017, Klatt2019}, $k_a = 0$, $k_p = 0$, $\epsilon = 1$; (A) the area model, $k_a = 1$, $k_p = 0$, $\epsilon = 0$; (CA) the centroid+area model, $k_a = 1$, $k_p = 0$, $\epsilon = 1$.
We present results for $p_0 = 3.4$, but have verified that results only qualitatively depend on this choice as long as $p_0 \leq 3.81$, where the system exhibits a solid response at small strains~\cite{bi2016motility,li2018role}.
Neglecting the Voronoi constraint on the shape of the particles, models AP and CA impose two constraints on each particles. 
Hence, the number of constraints equals the number of degrees of freedom, and these models are marginal. 
Conversely, models A and C are underconstrained, and hence are expected to have no mechanical rigidity.

We investigate the response to shear deformations of a $N=2000$ particles enclosed in a square box with length $L = \sqrt{N}$, using Lees Edwards periodic boundary conditions~\cite{lees1972computer}. 
We construct the Voronoi tessellations via the C++ Boost Voronoi library~\cite{schaling2011boost}, that accurately handles degenerate vertices.
All simulations start from initial configurations generated via the minimization of the energy of a set of points randomly placed in the simulation box.

We study the response to shear deformation in the limit of zero shear rate and zero temperature by resorting to athermal quasi-static shear simulations (AQS). 
These simulations involve the iteration of a two-step loop, a) increase of the shear strain, $\gamma \to \gamma +\delta\gamma$; b) relaxation of the energy of the system.
For computational efficiency, the relaxation of the system is obtained via an energy minimization algorithm rather than resorting to a damped dynamics. 
In our study, we fix $d\gamma = 10^{-3}$ and resort to the conjugate gradient minimization algorithm as implemented in the GSL library~\cite{galassi2018scientific}.

In AQS simulations, the dependence of the elastic energy on the shear strain comprises a series of continuous elastic branches, within which the energy increases, and is punctuated by sudden drops corresponding to failure events.
We identify these events via a standard thresholding procedure on the energy changes in a strain increment. 
Due to the many-body nature of the interaction potential, the stress is not directly accessible via the virial theorem~\cite{zimmerman2004calculation}.
Hence, here we evaluate the stress as $\sigma = \frac{dU}{d\gamma}$, approximating the elastic energy $U$ with a polynomial within each elastic branch, to reduce numerical noise.

\section{Mechanical response}
We illustrate the strain dependence of the energy and stress of the different models in figure~\ref{fig:energy_stress}(a).
At small strains, $\gamma \leq 3$, the AP model exhibits an elastic response followed by a plastic one, as common in molecular or colloidal systems, while the A and the C model are clearly unjammed.
The CA model appears to have a small but finite stress - in agreement with the expectation that, due to constraint counting~\cite{sussman2018no}, this model does not have a true jamming transition phase.

At larger strains, the energy of all models suddenly increases.
This increase marks the beginning of a crossover regime in the rheological properties (grey shaded area).
In the AP model, during this crossover the stress drops.
In the other models, this crossover entails a shear-jamming transition~\cite{Bi2011,Ciamarra2011} leading to the growth of the shear stress, followed by a stress drop.
In all cases, we assume the crossover ends when the shear stress starts fluctuating around zero. 
Due to the absence of mechanical rigidity of the final states we interpret the observed scenario as a strain weakening crossover.
The strain weakening crossover correlates with changes in the dynamical and geometric properties of the system.
Since the system is not in a steady-state, we evaluate the dynamics through a $\gamma$-dependent diffusivity~\cite{Chattoraj2011},
\begin{equation}
D_r(\gamma) = \langle (r(\gamma + d\gamma) - r(\gamma - d\gamma))^2\rangle/(2d\gamma),
\end{equation}
where $r = x-y\gamma$ or $r=y$ is a component of the not-affine displacement field and $d\gamma = 0.2$ is the scale of observation.
Figure~\ref{fig:energy_stress}(b) shows that the diffusivity sharply drops at the beginning of the crossover in models A,C, which undergo a shear-jamming transition, and of the AC model, which also rigidifies.
In all models, the diffusivity reaches minute values at the end of the crossover, $\simeq 10^{-4}$.
This drop in the diffusivity is the proxy of the emergence of an almost affine response of the system to shear deformations: only after strain increments of order $10^4$ the (non-affine) separation between close particles varies of a quantity comparable to the typical inter-particle distance.

\section{Evolution of geometrical properties}
We investigate the evolution of geometrical properties of the system associating to each particle an eccentricity $e = \sqrt{1-\left(\lambda_2/\lambda_1\right)^2}$, where $\lambda_1 \ge \lambda_2$ are the eigenvalues of the covariance matrix of its vertices~\cite{Li2021}.
The direction of the eigenvector associated with $\lambda_1$ defines the cell's polarity, ${\bf p} = {\bf u}_1/{\mid{\bf u}_1\mid}$.
The nematic order parameter is then $S = \langle 2 \cos^2{\phi_i} -1 \rangle$, $\phi_i$ being the angle between the polarity of cell $i$ and the nematic director.
The average eccentricity of the cells grows at small strains, but then slightly drops within the strain weakening crossover, as apparent from figure~\ref{fig:energy_stress}(c) (red).
The polarities of the cells remain roughly uncorrelated, as the nematic order parameter (blue, right axis) does not vary significantly, if not during the initial transient of the AP model.
Hence, in the strain weakened regime cells are more elongated than in the initial unstrained configuration, but their polarities are not aligned over long distances.
\begin{figure}[t!]
	\centering
	\includegraphics[width = 8cm]{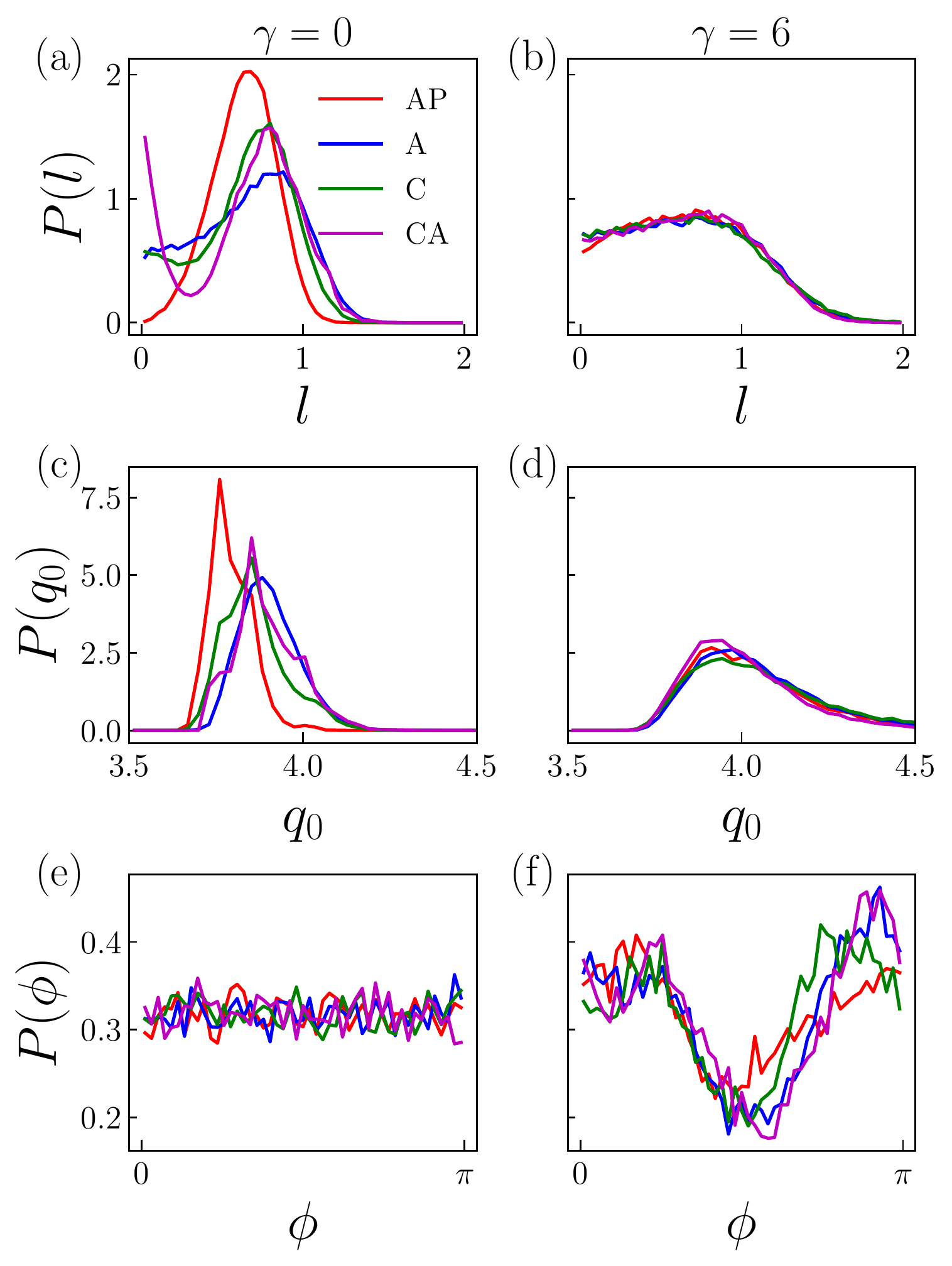}
	\caption{Distribution of
		cell edge-lengths $l$ (a,b), 
		shape index $q_0$ (c,d) and orientation $\phi$ (e,f), for $\gamma=0$ and $\gamma=6$ respectively.
	}
	\label{fig:dist_q_l_theta}
\end{figure}

In the strain weakened state the system rearranges easily and hence possesses an extensive number of soft spots~\cite{lemaitre2007plastic, Kim2018, li2018role, popovic2020inferring, Li2021,karmakar2010statistical,lin2014density}, structural rearrangements induced by a small increment $s$ of the local shear stress.
Accordingly, the distribution $P(s)$ of the local stress increments needed to trigger a rearrangement scales as $s^\theta$ with $\theta > 0$ in the solid phase, and $\theta = 0$ in the fluid one.
It is generally arduous to infer $P(s)$ from a snapshot of the system, and hence to correlate geometrical and mechanical properties. 
However, when the energy of the particles is dominated by their perimeter, the stress needed to trigger a T1 transition leading to the disappearance of an edge of length $l$ is $s \propto l$. 
Hence, $P(s) \propto P(l)$~\cite{popovic2020inferring,krajnc2018fluidization}.
For the parameter values we have considered, this occurs in the AP model~\cite{Bi, Kim2018, li2018role, popovic2020inferring, Li2021}.
Indeed, figures~\ref{fig:dist_q_l_theta}(a) and (b) reveal that the edge-length distribution of the AP model quickly vanishes as $l \to 0$ in the solid phase at small $\gamma$, while it saturates to a finite value in the strain weakened regime, implying $\theta = 0$.
For the other models, $P(l)$ is not in principle informative of the degree of mechanical stability, and we find $P(l = 0) > 0$ at all $\gamma$. 
Surprisingly, however, we find that $P(l)$, which is model dependent at $\gamma = 0$, becomes model independent in the strain weakened regime, as in figures~\ref{fig:dist_q_l_theta}(a,b).
\begin{figure}[t!]
	\centering
	\includegraphics[width= 0.48\textwidth]{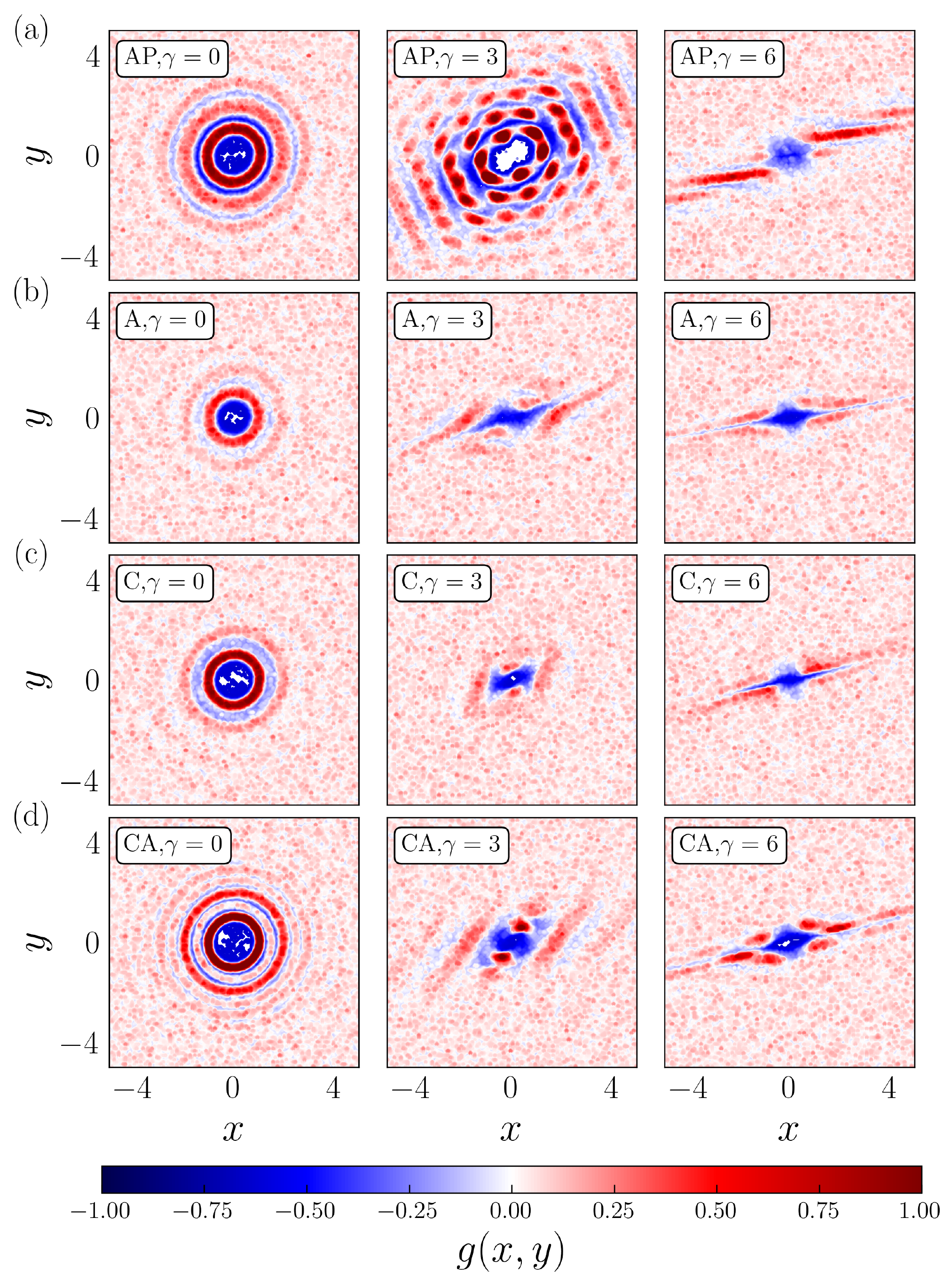}
	\caption{Density correlation function $g(x,y)$, for (a) AP, (b) A, (c) C and (d) CA models, respectively. 
		The different columns refer to $\gamma = 0,3$ and $6$, from left to right.
		\label{fig:snap}
	}
\end{figure}

Other geometrical properties also appear to become model-independent in the strain weakened regime. 
For instance, figures~\ref{fig:dist_q_l_theta}(c) and (d) show that the distribution of the shape index of the particles\cite{park2015unjamming,Bi}, as set by their actual shape $q_0 = \frac{p_i}{\sqrt a_i}$, is model dependent in the initial configuration and becomes model independent in the asymptotic strain weakened regime. 
The average $q_0$ value increases as the system is strained, consistent with the increase of the average particle eccentricity.
The distribution of the orientations $\phi$ of the polarity vectors of the cells, ${\bf p} = (\cos\phi,\sin\phi)$, which is flat in the initial configuration (e), is model dependent at intermediate strains (not shown) and becomes approximately model independent in the strain weakened regimes (f).
In this regime, the distribution has a well pronounced minimum at around $\phi \simeq \pi/2$, and for all but the AP model, two asymmetric maxima at $\phi \simeq \pi/4$ and $3\pi/4$.
The results of figure~\ref{fig:dist_q_l_theta} thus indicate that in the strain weakened regime particles acquire similar model-independent typical shapes, and that they are similarly oriented in space.

The above analysis suggests that all models acquire a same typical structure in the strain weakened regime.
\begin{figure}[t!]
	\centering
	\includegraphics[width= 0.48\textwidth]{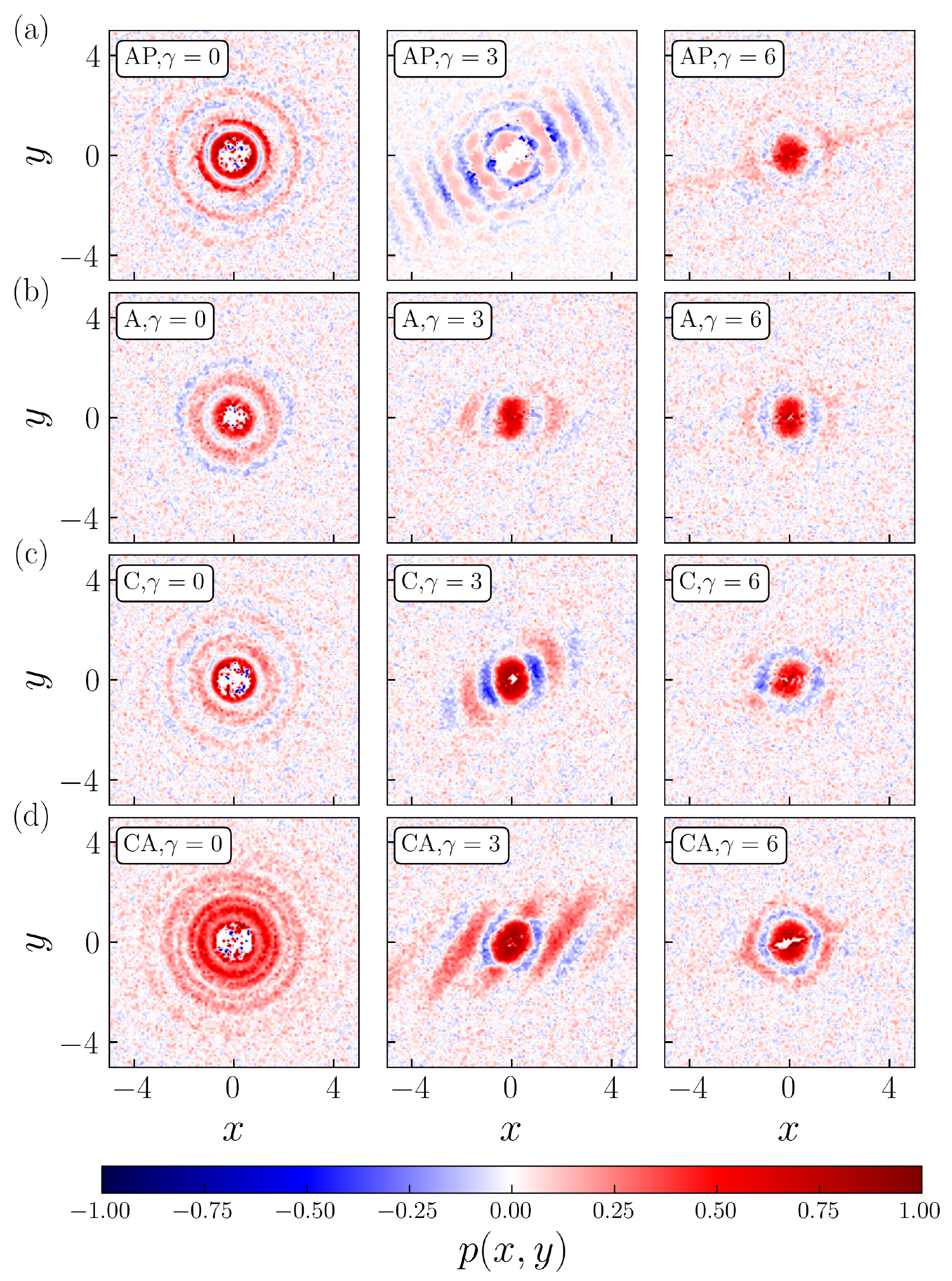}
	\caption{Evolution of 2d polar correlation under strain for (a) AP, (b) A, (c) C and (d) CA model respectively. The different columns refer to $\gamma = 0,3$ and $6$, from left to right.
		\label{fig:snap_pc}
	}
\end{figure}

\begin{figure}[t!]
	\centering
	\includegraphics[width = 8cm]{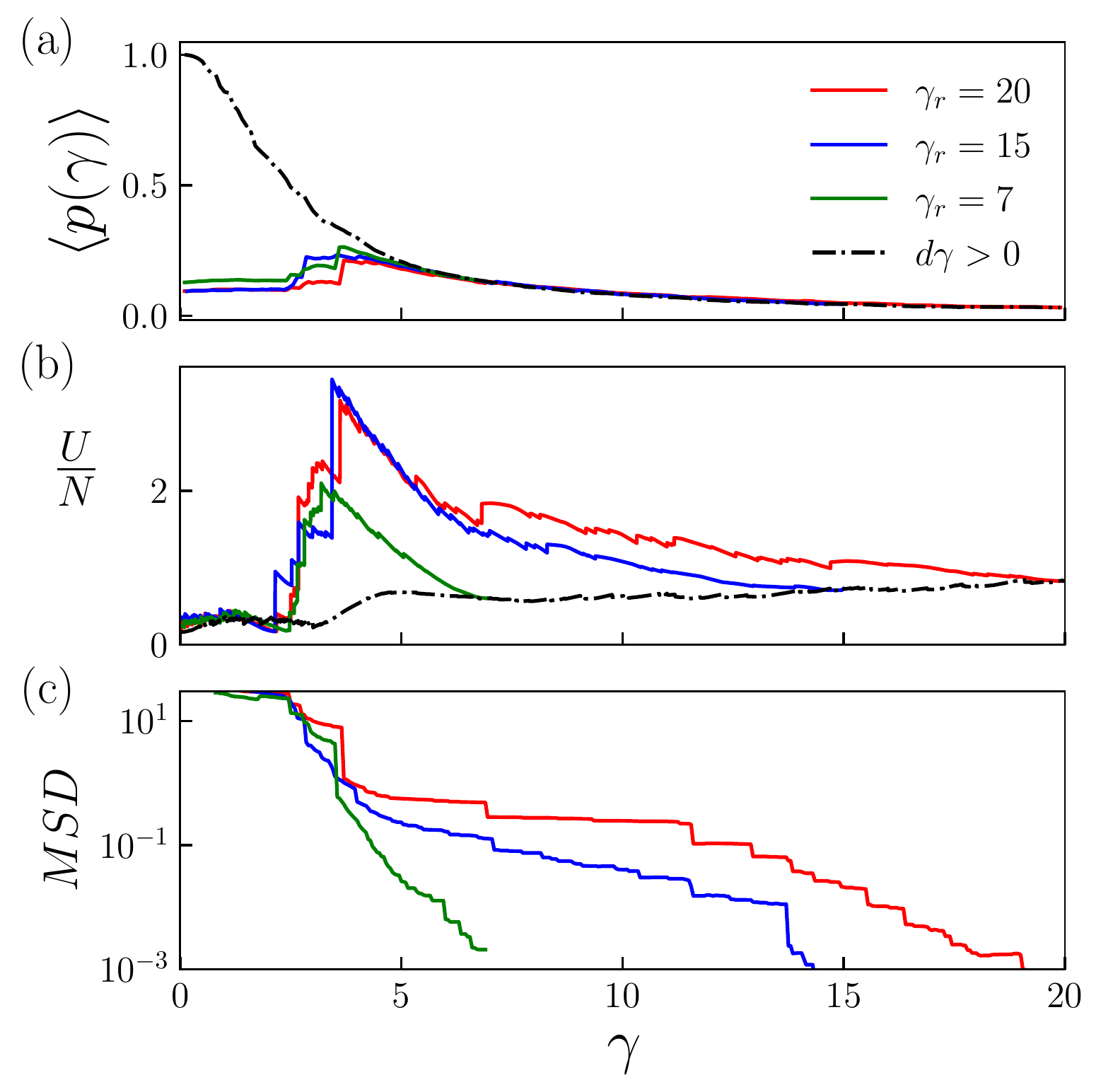}
	\caption{The dashed-dot line illustrates the evolution of the persistence (a) and of energy (b) of the AP model, as the strain slowly increases. 
		The other lines show how these quantities evolve as the strain is reversed, after having reached a maximum value $\gamma_r$. 
		Panel (c) illustrates the increase of the mean square displacement as the strain decreases.
		At large $\gamma$ in the strain weakened regime, the persistence is reversible while the energy is not.
	}
	\label{fig:reverse_en}
\end{figure}

To test this hypothesis, we investigate the 2d density correlation function, $g(x,y)$, and the 2d correlation function of the cell's polarities,
\begin{equation}
p(x,y) = \langle \cos 2(\phi_i - \phi_j) \rangle
\end{equation}
where the average is over all cell pairs with a distance ($x$, $y$).
In initial configurations of all models, $g(x,y)$ is radially symmetric, with correlations more pronounced in the AP and CA models, as in the left most column of figure~\ref{fig:snap} and figure~\ref{fig:snap_pc}.
As the strain increases, the evolution of the structure is strongly model dependent.
For instance, figure~\ref{fig:snap}(middle column) clarifies that only the AP model transiently develops crystalline order over an appreciable length scale, the underlying crystal corresponding to an obliquely strained triangular lattice, as apparent from the location of the first six peaks.
However, strong similarities between the correlation functions of the different models emerge in the strain weakened regime (right column).
In particular, a four pointed star region with $g \simeq -1$, is present in all models.
The star is slightly rotated with respect to the $x$-axis, and the two almost horizontal arms are longer than the vertical ones. 
Considering that square-lattice of central-force model are generally observed to be mechanically unstable~\cite{PhysRevB.43.3450,Mao2015} at zero temperature, we speculate that the emergence of this short-ranged tetratic order~\cite{donev2006tetratic,hou2020emergent,geng2009theory}, is a signature of the existence of an extensive number of soft spots that allow the system to flow exerting minute mechanical stress.

\begin{figure*}[!t]
	\centering
	\includegraphics[width = 0.9\textwidth]{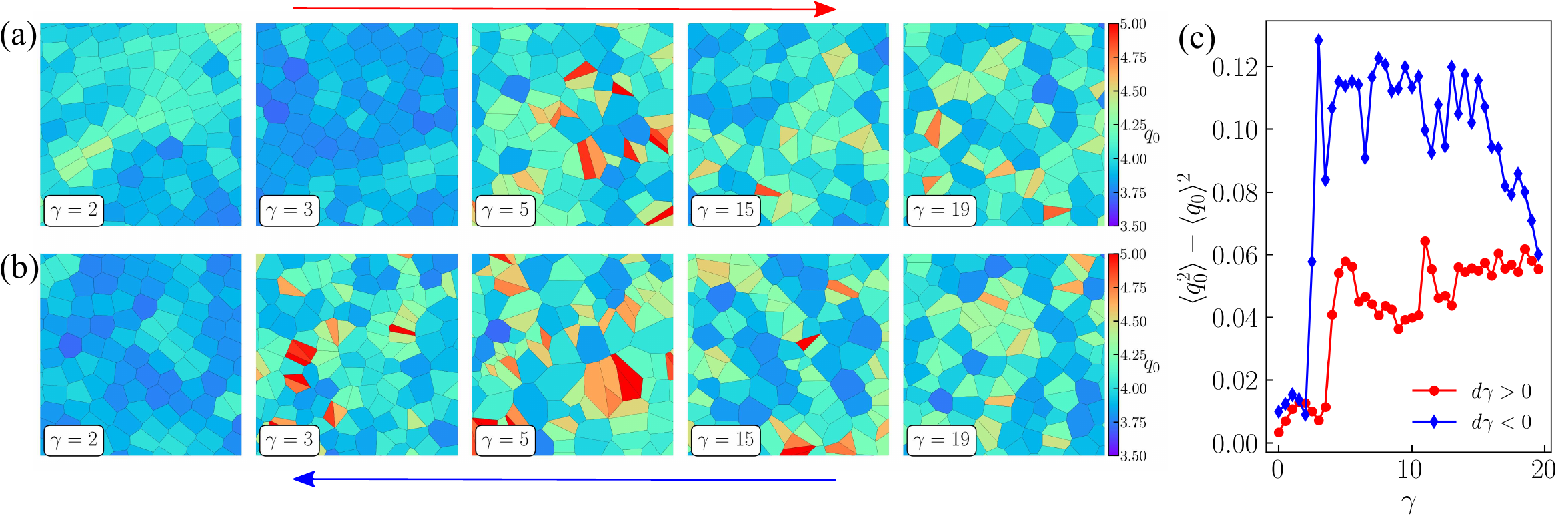}
	\caption{Snapshots of a small region of configurations of the AP model reached increasing (a) and decreasing (b) the strain. Particles are colour coded according to the value of their shape index, $q_0$, and the shear is reversed at $\gamma_r = 20$. 
		Panel (c) illustrates the $\gamma$ dependence of the fluctuations of $q_0$. 
	}
	\label{fig:reverse_area_snap}
\end{figure*}

\section{Memory in the strain weakened regime}
In the strain weakened regime, the system is jammed: the energy is finite and the shear stress fluctuates about minute values (see figure~\ref{fig:energy_stress}). 
In addition, the diffusivity is strongly suppressed but not fully inhibited as it occurs, e.g., in a strain induced diffusive-absorbing phase transition of colloidal systems~\cite{GadalaMaria1980,Tsai2003,Grebenkov2008,Zheng2021}.
This suggests that cell-cell contacts survive large strains and that the contact network keeps memory of its initial state.
To check this hypothesis, we define the persistence $p_i(\gamma)$ of particle $i$ as the fraction of Voronoi neighbors of the $\gamma = 0$ configuration which are also Voronoi neighbors at strain $\gamma$, and study the average persistence as the strain first increases up to $\gamma_r$, and then decreases to zero.
Figure~\ref{fig:reverse_en} (a) shows that the persistence of the AP model behaves reversibly, at large $\gamma$ in the strain weakened regime, regardless of the value of $\gamma_r$. 
The other models behave similarly.
Since the persistence is a measure of the evolution of the topology of the system, we understand that as the strain is reduced the system gradually recovers previously broken contacts.
This finding indicates that in the weakened regime the system acts as a memory device. 
A scalar $x$ can be recorded by straining a weakened configuration up to $\gamma \propto x$.
To read the written quantity, one should strain the system in reverse until the topology on the starting configuration is recovered.
We expect the maximum strain the system is able to record to be of order $D_y^{-1}\simeq 10^4$ - a too large value to be tested numerically.

While the observed phenomenology appears analogous to the memory through path reversal~\cite{Keim2019} which characterizes, e.g., highly viscous liquids~\cite{Taylor1986}, there are interesting and crucial differences.
Indeed, in viscous liquids memory emerges as the microscopic dynamics is reversible under time reversal.
This cannot occur here as, in the strain weakened regime, stress avalanches indicative of irreversible dynamics are clearly present.
To demonstrate this point, we investigate in
figure~\ref{fig:reverse_en}(b) the evolution of the energy under strain reversal.
The energy slowly increases with $\gamma$ in the strain weakened regime, but it does not decrease when the strain is reversed.
Instead, the energy increases up to a strain $\gamma_t$ along the  reverse path, where the system resets itself through a gigantic failure event to a low energy random state that is similar to the unstrained one.
Similarly, we observe in figure~\ref{fig:reverse_en} (c) the mean square displacement $\langle (\bf{r}(\gamma)-\bf{r}(\gamma_r))^2 \rangle$ also increases as $\gamma$ decreases.
This behavior does not depend on the strain $\gamma_r$ reached before reversing the strain increment.

To further illustrate that cell's properties change during the reverse shear, we provide in figure~\ref{fig:reverse_area_snap} partial snapshots of the system at different $\gamma$ values reached as the strain increases (a) and decreases (b), coloring each cell according to the values of its own shape index, $q_0$.
Configurations with the same $\gamma$ value, reached on increasing or increasing $\gamma$, clearly differ. 
These differences are quantified in panel (c), which illustrates that configurations reached as the strain decreases have much larger $q_0$ fluctuations.

\section{Conclusion}
Our results demonstrate that, at high density, the correlation between energy and shape of the particles induces a previously unreported strain-weakening crossover in systems of soft deformable particles.
This involves a drop in the shear stress to minute values as the saturates and particles acquire model-independent shapes.
In the weakened regime, the contact network evolves reversibly, while the particles' positions do not.
This crossover is enabled by particles' deformability, that weakens the system by allowing for the emergence of short-ranged tetratic order.

We notice that Ref.~\cite{popovic2020inferring} investigated the AP model under shear, focusing on the Vertex model, and did not observe any weakening crossover. 
In the Vertex model the degrees of freedom are the vertices rather than the cell centers, and topological transitions are artificially triggered according to some rule based on a chosen threshold for the edge length.
Possibly, the presence of this rule might critically affect the response to shear, and indeed we observe edge-lengths of order $10^{-5}$, while the smallest edged reported in Ref.~\cite{popovic2020inferring} is of order $10^{-2}$. 
Alternatively, it is possible that the observed phenomenology stems from the Voronoi constraint shared by all our models, i.e. from the fact that the particles' shape are identified with the Voronoi cells.
It is not obvious that this should be the case, as the Voronoi constraint should stiffen the system, rather than promoting its weakening~\cite{sussman2018no}.

So far, our reported strain weakening crossover has not been observed in experiments.
For instance, when subjected to an imposed strain, foams do not weaken homogeneously as we observe here, but rather develop a shear band~\cite{Debregeas2001,Kabla2003}. 
There could exist systems exhibiting a strain weakening crossover.
The response of epithelial tissue to shear strain deformations, for instance, has not been experimentally investigated. Similarly, the response of high density microgel suspensions in a 2d-geometry is unexplored.
Hence, it would be interesting to experimentally ascertain the rheological properties of these systems, and of related systems of soft deformable particles at high density.

\section*{Conflicts of interest}
There are no conflicts to declare.

\section*{Acknowledgements}
We thank the Singapore Ministry of Education through the Academic Research Fund 2019-T1-001-032 (Tier 1). 
The computational work for this article was fully performed on resources of the National Supercomputing Centre, Singapore.




\bibliography{citations} 
\bibliographystyle{rsc.bst} 

\end{document}